\documentstyle[aps,preprint,epsfig,floats]{revtex}

\def\slash#1{#1\!\!\!/}
\def\eqref#1{Eq.\ (\ref{#1})}

\begin{document}
\setcounter{page}{0}
\def\footnoterule{\kern-3pt \hrule width\hsize \kern3pt}
\tighten

\title{A Diagrammatic Approach to Crystalline Color Superconductivity}

\author{Jeffrey~A.~Bowers\footnote{Email address: {\tt jbowers@mit.edu}},
Joydip~Kundu\footnote{Email address: {\tt kundu@mit.edu}},
Krishna~Rajagopal\footnote{Email address: {\tt krishna@ctp.mit.edu}},
Eugene~Shuster\footnote{Email address: {\tt eugeneus@mit.edu}}}

\address{Center for Theoretical Physics \\
Massachusetts Institute of Technology \\
Cambridge, MA 02139 \\
{~}}

\date{MIT-CTP-3071,~  hep-ph/0101067,~ January 8, 2001.}
\maketitle

\thispagestyle{empty}

\begin{abstract}
We present a derivation of the gap equation for the crystalline
color superconducting phase of QCD which begins from a 
one-loop Schwinger-Dyson
equation written using a Nambu-Gorkov propagator
modified to describe the spatially varying condensate.
Some aspects of previous variational calculations 
become more straightforward when rephrased
beginning from a diagrammatic starting point.  This derivation 
also provides a natural base from which to generalize the
analysis to include quark masses, nontrivial crystal structures, 
gluon propagation at asymptotic densities, and nonzero temperature.  
In this paper, we analyze the effects of nonzero temperature on
the crystalline color superconducting phase.
\end{abstract}

\vfill\eject

\section{Introduction and Summary}

It is becoming widely accepted that at asymptotic densities
the ground state of QCD with three massless quarks is the
color-flavor locked (CFL) phase \cite{CFL,OtherCFL,ioffe}.
This phase features a condensate of Cooper pairs of
quarks which includes $ud$, $us$ and $ds$ pairs.  
The CFL phase persists for finite masses,
and even for unequal masses, so long as the 
differences are not too large \cite{ABR2+1,SW2}.
It also persists in the presence of a nonzero electron chemical
potential $\mu_e$, so long as 
$\mu_e$ is not too large \cite{rigidity}.  
In the absence of any interaction (and thus in the 
absence of CFL pairing) either a quark mass difference
or a nonzero $\mu_e$ pushes the Fermi momenta for different
flavors apart, yielding different number densities for
different flavors.  In the CFL phase, however, the fact that
the pairing energy is maximized when $u$, $d$ and $s$ 
number densities are equal
enforces this equality \cite{rigidity}.  This means
that if one imagines increasing either the strange quark
mass $m_s$ or $\mu_e$, nothing happens until a first
order phase transition, at which CFL pairing is disrupted,
(some) quark number densities
spring free under the accumulated tension, and a less
symmetric state of quark matter is obtained \cite{rigidity}.

We can study much of 
the physics of interest by focussing just on
pairing between massless up and down quarks with chemical
potentials
\begin{eqnarray}\label{mubardmu}
\mu_u&=&\bar\mu-\delta\mu\nonumber\\
\mu_d&=&\bar\mu+\delta\mu\ .
\end{eqnarray}
For $0<\delta\mu<\delta\mu_1$, the ground state 
is precisely that obtained for 
$\delta\mu =0$ \cite{Clogston,Bedaque,BowersLOFF}.  In this state,
red and green up and down quarks pair, yielding four quasiparticles
with superconducting gap $\Delta_0$ \cite{Barrois,BailinLove,ARW1,RappETC}.
And, the number density of red and green up quarks is the
same as that of red and green down quarks. 
As $\delta\mu$ is increased from zero, this BCS state
remains unchanged (and favored) 
because maintaining coincident Fermi surfaces 
maximizes the pairing
and thus the gain in interaction energy.  
As $\delta\mu$ is increased
further, the BCS state remains the ground state of
the system only as long as its negative interaction
energy offsets the large positive free energy cost 
associated with forcing the Fermi seas to deviate from their
normal state distributions. In the weak coupling limit, in
which $\Delta_0/\bar\mu\ll 1$, the BCS state persists while
$\delta\mu<\delta\mu_1=\Delta_0/\sqrt{2}$ \cite{Clogston,BowersLOFF}.

For $\delta\mu\gg\delta\mu_1$, only very weak pairing between like-flavor
quarks is possible \cite{Schaefer1Flav}. Near the unpairing
transition, however, another phase intervenes. This is the ``LOFF''
state, first explored by Larkin and Ovchinnikov and Fulde
and Ferrell in the context of electron superconductivity
in the presence of magnetic impurities \cite{LO,FF}. 
Translating LOFF's results to the case of interest, 
the authors of Ref. \cite{BowersLOFF}
found that for $\delta\mu\gtrsim\delta\mu_1$ it is
favorable to form a state in which the $u$ and $d$ Fermi
momenta are given by $\mu_u$ and $\mu_d$  as in the absence
of interactions, and are thus not equal, but pairing
nevertheless occurs.  Whereas in the BCS state, obtained
for $\delta\mu<\delta\mu_1$, pairing occurs between
quarks with equal and opposite momenta, when $\delta\mu\gtrsim\delta\mu_1$
it is favorable to form a condensate of Cooper pairs with nonzero
total momentum.  This is favored because 
pairing quarks with momenta which
are not equal and opposite gives rise to a region of phase space
where each of the two quarks in a Cooper pair can be
close to its Fermi surface, even when the up and
down Fermi momenta differ, and such pairs can
be created at low cost in free energy.\footnote{LOFF
condenates have also recently been considered in two other contexts. 
In QCD with $\mu_u<0$, $\mu_d>0$ and $\mu_u=-\mu_d$, one has
equal Fermi momenta for $\bar u$ antiquarks and $d$ quarks,
BCS pairing between them, and consequently a 
$\langle \bar u d\rangle$ condensate \cite{SonStephIsospin}.  
If $-\mu_u$ and $\mu_d$ differ,
and if the difference lies in the appropriate range, a LOFF
phase with a spatially varying $\langle \bar u d\rangle$
condensate results \cite{SonStephIsospin}. 
Suitably isospin asymmetric nuclear matter may also admit LOFF pairing, as 
discussed recently in Ref. \cite{Sedrakian}.}
Condensates of this sort spontaneously break
translational and rotational invariance, leading to gaps which
vary periodically in a crystalline pattern.  If in some
shell within the quark matter core of a neutron star
(or within a strange quark star) the quark chemical potentials
are such that crystalline color superconductivity
arises, as occurs for a wide range of reasonable parameter
values,
rotational vortices may be pinned in this shell, making
it a locus for glitch formation \cite{BowersLOFF}. 
Rough estimates of the pinning force suggest that it 
is comparable to that for a rotational vortex pinned
in the inner crust of a conventional neutron star,
and thus may yield glitches of phenomenological interest \cite{BowersLOFF}.

The authors of Ref. \cite{BowersLOFF} studied crystalline
color superconductivity in a simplified model 
with two flavors of quarks with chemical potentials (\ref{mubardmu})
which interact via a four-fermion interaction with
the quantum numbers of single gluon exchange. In the LOFF state,
each Cooper pair has total momentum $2{\bf q}$ with 
$|{\bf q}|\approx 1.2\delta\mu$. The direction of
${\bf q}$ is chosen spontaneously. The LOFF
phase is characterized by a gap parameter $\Delta$ and
a diquark condensate, but not by an energy gap: the quasiparticle
dispersion relations vary with the direction of the momentum,
yielding gaps which vary from zero up to a maximum of $\Delta$.
The condensate is dominated by those regions in momentum
space in which a quark pair with total momentum $2{\bf q}$
has both members of the pair within $\sim \Delta$ of their respective
Fermi surfaces. The gap equation which determines $\Delta$
was derived in Ref. \cite{BowersLOFF} using variational
methods, along the lines of Refs. \cite{FF,Takada2}.
This gap equation can then be used to show that crystalline
color superconductivity is favored over no $ud$ pairing
for $\delta\mu<\delta\mu_2$.  Here, $\delta\mu_2\approx 0.754 \Delta_0$
if the coupling is weak \cite{LO,FF,Takada2}
and if there is no
interaction in angular momentum $J=1$ 
channels \cite{BowersLOFF}. For stronger coupling
and for varying choices of interaction, $\delta\mu_2$ 
changes \cite{BowersLOFF}.

Crystalline color superconductivity is favored for 
$\delta\mu_1<\delta\mu<\delta\mu_2$. As $\delta\mu$ increases,
one finds a first order phase transition from the ordinary
BCS phase to the crystalline color superconducting phase
at $\delta\mu=\delta\mu_1$ and then a second order
phase transition at $\delta\mu=\delta\mu_2$ at which $\Delta$          
decreases to zero.
Analysis of the Ginzburg-Landau effective potential which describes
physics near  $\delta\mu_2$ shows that  
$\Delta\sim(\delta\mu_2-\delta\mu)^{1/2}$ 
for $\delta\mu\rightarrow\delta\mu_2$ \cite{BowersLOFF}.
Because the condensation
energy in the LOFF phase is much smaller than that of the BCS condensate
at $\delta\mu=0$, the value of $\delta\mu_1$ is almost identical
to that at which the naive unpairing transition from the 
BCS state to the state with no pairing would occur if
one ignored the possibility of a LOFF phase.  For all practical
purposes, the LOFF gap equation is not required in order
to determine $\delta\mu_1$. The LOFF gap equation
is used to 
determine $\delta\mu_2$
and the properties of the LOFF phase \cite{BowersLOFF}. 
For example, it determines the coefficients in
the Ginzburg-Landau effective 
potential valid
near $\delta\mu_2$ \cite{Takada2,BowersLOFF}.

The variational derivation of the gap equation for the crystalline
color superconducting phase is somewhat cumbersome \cite{BowersLOFF}.
One constructs a variational ansatz in which only quarks within
a ``pairing region'' are allowed to pair, minimizes the
free energy with respect to all variational parameters (two per mode
in momentum, color, flavor and spin space), and obtains
a self-consistency relation which may then be solved to obtain
$\Delta$.  The intricacy arises from the fact that the definition
of the boundary of the pairing region involves $\Delta$ itself.
A derivation in which one simply makes an ansatz for the 
quantum numbers of the condensate and then ``turns a field-theoretical
crank'' and sees this intricate result emerge
would be helpful both by virtue of being more straightforward
and because the use of variational methods to obtain a
gap equation is by now less familiar to many readers.
We provide such a diagrammatic derivation here.

Furthermore, and as we explain at appropriate points
in our presentation of the derivation in Sections II, III and IV, many 
generalizations are amenable to analysis using the 
formalism we present here:
\begin{itemize}
\item
In Section V, we include the effects of nonzero temperature.
We calculate the critical temperature $T_c$ above
which the crystalline color superconducting condensate
vanishes and show that for $\delta\mu\rightarrow\delta\mu_2$,
$T_c\rightarrow 0.39\Delta$, as previously known \cite{Takada2}.
\item
It should also be straightforward to generalize the analysis
to include three flavors of quarks with differing masses,
and thus to study the crystalline color superconducting
phases expected where either $\mu_e$ or $m_s$ is just larger 
than that at which the CFL phase is lost \cite{rigidity}.
This, not the toy model we analyze here, is the case
of physical interest.  
\item
As in Ref. \cite{BowersLOFF}, we restrict our attention here to the 
simplest possible ``crystal'' structure, namely that in which
the condensate varies like a plane wave. Wherever
this condensate is favored over the homogeneous BCS condensate
and over the state with no pairing at all (i.e. 
where $\delta\mu_1<\delta\mu<\delta\mu_2$) we expect that the
true ground state of the system is a condensate which
varies in space with some more complicated spatial dependence.
The formalism we set up can be generalized to derive gap
equations for, and hence to analyze and compare, condensates
with arbitrary crystal structures in order to learn which one is 
favored.
\item
The diagrammatic analysis we present uses a point-like interaction
between quarks, but the formalism is easily generalized to treat
the exchange of a propagating gluon, as appropriate at
asymptotically high densities. Even if such analyses,
pioneered at $\delta\mu=0$ in Ref. \cite{Son} and since
studied in considerable detail by many authors \cite{ioffe},
are to date of quantitative value only at inaccessibly
high densities \cite{Shuster}, it would be very interesting
to see crystallization occurring in a controlled analysis
beginning directly from the QCD Lagrangian.
\end{itemize}

\section{The Gap Equation for Crystalline Color Superconductivity}

In the ordinary BCS phase, pairing between quarks 
with momentum ${\bf p}$ and $-{\bf p}$ is described
in the standard Nambu-Gorkov formalism by introducing
an eight-component field
\begin{equation} \label{PsiBCS}
\Psi(p) = \left(\begin{array}{l} \psi(p) \\
\bar\psi^T(-p) \end{array}\right)\ ,
\end{equation}
such that, in this basis, the inverse quark propagator takes the form
\begin{equation} \label{SinvBCS}
S^{-1}(p) = \left[\begin{array}{cc} \slash{p}+\mu\gamma_0
& \bar{\bf \Delta}(p) \\ {\bf \Delta}(p) & 
(\slash{p}-\mu\gamma_0)^T \end{array}\right]\ .
\end{equation}
Here, $\bar{\bf \Delta} = \gamma_0 {\bf \Delta}^{\dagger} \gamma_0$
and $\bar{\bf \Delta}$ is a matrix with
color, flavor and Dirac indices which have all been suppressed.
The diagonal blocks 
correspond to ordinary propagation
and the off-diagonal blocks reflect the possibility of
``anomalous propagation'' in the presence of a diquark
condensate 
$\langle \psi({\bf x}) \psi({\bf x})\rangle\propto {\bf \Delta}$.

In the crystalline color superconducting phase\cite{LO,FF,BowersLOFF}, 
the condensate is made up of pairs of $u$ and $d$ 
quarks with momenta such that the total momentum
of each Cooper pair is given by $2{\bf q}$
with $|{\bf q}|\approx 1.2 \delta\mu$.  The direction of 
${\bf q}$ is chosen spontaneously.   Such a condensate
varies periodically in space, with wavelength $\pi/|{\bf q}|$:
\begin{equation}\label{LOFFcondensate}
\langle \psi({\bf x}) \psi({\bf x})\rangle\propto {\bf \Delta} 
e^{2 i {\bf q}\cdot {\bf x}}\ .
\end{equation}
The spatial dependence (\ref{LOFFcondensate}) is only the simplest
possible choice.  Wherever (\ref{LOFFcondensate}) is favored over both the
ordinary BCS state and the state with no pairing at all,
we expect that the true ground state of the system
will include Cooper pairs with their 
respective $|{\bf q}|$'s taking on the same, energetically favored, value,
but choosing one of several spontaneously selected directions.  
The result would be a condensate which varies in
space like a sum of plane waves. For example, a cubic crystal
arises as a sum of six plane waves.  
The favored crystal structure
for the crystalline color superconductor
is not known.  In this paper, as in Ref. \cite{BowersLOFF},
we only consider the simplest possibility (\ref{LOFFcondensate}).

Although we expect that crystalline color superconductivity
occurs whenever the mass difference or chemical potential
difference between any two flavors of quarks is just larger
than the maximum value which the standard BCS state can
tolerate, for concreteness we shall follow Ref. \cite{BowersLOFF}
and only consider pairing between massless $u$ and $d$ quarks
with chemical potentials (\ref{mubardmu}).
In the condensate (\ref{LOFFcondensate}),
$u$ quarks with momentum
${\bf p}+{\bf q}$ pair with $d$ quarks with momentum $-{\bf p}+{\bf q}$.  
To describe this, we must use a modified Nambu-Gorkov spinor
defined as 
\begin{equation} \label{Psi}
\Psi(p,q) = \left(\begin{array}{l} \psi_u(p+q) \\ \psi_d(p-q) \\
\bar\psi^T_d(-p+q) \\ \bar\psi^T_u(-p-q) \end{array}\right)\ .
\end{equation}
Note that flavor indices are now explicit, which will be convenient
below.  The central change we have made in going from 
(\ref{PsiBCS}) to (\ref{Psi}) is
to modify the momentum dependence.  
Note that by $q$ we
mean the four-vector $(0,{\bf q})$.  The Cooper pairs
have nonzero total momentum, but the ground state condensate 
(\ref{LOFFcondensate}) is static.
The change from (\ref{PsiBCS}) to (\ref{Psi})
can be seen as a change of basis.  In the presence of 
a crystalline color superconducting condensate, anomalous
propagation does not only mean picking up or losing two quarks
from the condensate. It also means picking up or losing momentum $2 {\bf q}$.
If we tried to describe this using the original basis
(\ref{PsiBCS}), the inverse quark propagator would no longer
be diagonal in momentum space.  The new basis (\ref{Psi}) has 
been chosen so that the inverse quark propagator 
in the crystalline color superconducting phase is diagonal
in $p$-space and is given by
\begin{equation} \label{Sinv}
S^{-1}(p,q) = \left[\begin{array}{cccc} \slash{p}+\slash{q}+\mu_u \gamma_0
& 0 & -\bar{\bf \Delta}(p,-q) & 0 \\ 0 & \slash{p}-\slash{q}+\mu_d
\gamma_0 & 0 & \bar{\bf \Delta}(p,q) \\ -{\bf \Delta}(p,-q) & 0 &
(\slash{p}-\slash{q}-\mu_d \gamma_0)^T & 0 \\ 0 & {\bf \Delta}(p,q) 
& 0 & (\slash{p}+\slash{q}-\mu_u \gamma_0)^T \end{array}\right]\ .
\end{equation}
$2{\bf p}$ 
is the relative momentum of the quarks in a given pair, and
is different for different pairs. 
In the gap equation below, we shall
integrate over $p_0$ and ${\bf p}$, as we sum the contribution
of all pairs.  $2{\bf q}$ is the center
of mass momentum of every pair in the condensate; it is
a constant and
thus will not be integrated over.  It is convenient to denote
flavor indices explicitly in (\ref{Sinv}) because 
we are describing the situation where $\mu_u\neq\mu_d$.  It is
straightforward to introduce different quark masses in \eqref{Sinv},
but then the calculations become more involved and we
therefore defer this to a future publication. 
Note that the condensate is explicitly antisymmetric
in flavor. Color and Dirac indices remain suppressed.  As desired,
the off-diagonal blocks describe anomalous propagation in
the presence of a condensate of diquarks with momentum $2{\bf q}$.
The choice of basis we have made is  analogous to
that introduced previously in the analysis
of a crystalline quark-antiquark condensate\cite{chiralcrystal}.
This work also points the way toward the generalization
of (\ref{Psi}) needed to handle a condensate which
varies in space like a cubic crystal rather than the
plane wave (\ref{LOFFcondensate}).

We wish to obtain the gap by solving the one-loop Schwinger-Dyson
equation, given by
\begin{equation} \label{SDeq}
 S^{-1}(k,q)-S_0^{-1}(k,q) = i g^2 \int \frac{d^4p}{(2\pi)^4}
 \Gamma_\mu^A S(p,q)\Gamma_\nu^B D_{AB}^{\mu\nu}(k-p)\ .
\end{equation}
Here, $D_{AB}^{\mu\nu}$ is the gluon propagator, $S$ is the full quark
propagator, whose inverse is given by (\ref{Sinv}), and $S_0$ is the
fermion propagator in the absence of interaction, given by $S$ with
${\bf \Delta}=0$.  $S_0$ looks unusual, because it depends on
both $k$ and the ``offset'' ${\bf q}$. This is a consequence 
of our choice of basis (\ref{Psi}), and would be a legitimate
if perverse way to describe noninteracting fermions.  This
choice of basis is natural in the crystalline 
color superconducting phase. The
vertices are defined as follows: 
\begin{equation} \label{vertex}
\Gamma_\mu^A = \left(\begin{array}{cccc} \gamma_\mu\lambda^A/2 & 0 & 0
& 0 \\ 0 & \gamma_\mu\lambda^A/2 & 0 & 0 \\ 0 & 0 &
-(\gamma_\mu\lambda^A/2)^T & 0 \\ 0 & 0 & 0 &
-(\gamma_\mu\lambda^A/2)^T \end{array}\right) .
\end{equation}
Note that in \eqref{SDeq} we have chosen to work in Minkowski space. We
will continue to work in Minkowski space until we obtain the
gap equation itself, which we will then write in Euclidean
space for
computational convenience.

We defer the analysis of the crystalline color superconducting
phase at asymptotically high densities to future work. 
In this paper, as in Ref. \cite{BowersLOFF}, 
we choose to caricature the interaction
between quarks as a point-like four-fermion interaction
with the quantum numbers of single-gluon exchange.
This means that in (\ref{SDeq}), we make the replacement
\begin{equation}\label{pointlike}
g^2 D_{AB}^{\mu\nu}\rightarrow - 3G g^{\mu\nu}\delta_{AB}
\end{equation}
where $G$, normalized as in Ref. \cite{ioffe}, 
is a dimensionful coupling constant
which parametrizes the strength of the interaction
between quarks.  Reasonable choices for $G$, motivated by 
zero density hadron phenomenology, yield a BCS gap on the order of 
100 MeV at $\bar\mu=400$ MeV in the absence of any chemical
potential difference $\delta\mu$ \cite{ioffe}.

Once we have removed the gluon propagator from (\ref{SDeq}),
we see that the right-hand side is independent of $k$.  
The left-hand side must therefore be independent of $k$ as well,
meaning that ${\bf \Delta}$ in (\ref{Sinv}) must be independent
of $p$.  ${\bf \Delta}$ does depend on the common momentum
of all the Cooper pairs, $2{\bf q}$.
Choosing an ansatz for ${\bf \Delta}$ is straightforward
once we have understood that it must be independent of $p$.
Single gluon exchange is attractive in the color-antisymmetric
(${\bf \bar 3}$), flavor antisymmetric, Lorentz scalar
or pseudoscalar channels.  Instanton effects favor the scalar
condensate, and we therefore make the ansatz
\begin{equation}\label{simpleansatz}
{\bf \Delta}^{\alpha\beta}(p,q)=\epsilon^{\alpha\beta 3} C\gamma_5 
\Delta
\end{equation}
for the gap matrix, where  $\alpha$ and $\beta$ are color indices, running
from 1 to 3, and $C=i\gamma_0\gamma_2$. 
$\Delta$ has no remaining indices and all the matrix structure
has now been written explicitly. $\Delta$ does depend on $|{\bf q}|$,
although we do not denote this dependence explicitly, 
but does not depend on $p$ or the direction of ${\bf q}$.

After some algebra (essentially the determination of $S$ given
$S^{-1}$ specified above), and upon suitable projection,
the Schwinger-Dyson equation (\ref{SDeq}) reduces
to a gap equation for the gap parameter $\Delta$ given
(in Euclidean space) by
\begin{equation} \label{gapeqdelall}
\Delta = 2 G \int \frac{d^4p}{(2\pi)^4} \frac{4 \Delta w}{w^2 - 4
\left[ (|{\bf p}|^2 - (i p_0+\delta\mu)^2) (\bar{\mu}^2 - |{\bf q}|^2) + 
({\bf p}\cdot{\bf q} + \bar\mu (i p_0 + \delta\mu))^2 \right]}
\end{equation}
where $w=|{\bf p}|^2 - |{\bf q}|^2 - (i p_0+\delta\mu)^2 + \bar{\mu}^2
+ \Delta^2$.

We analyze the gap equation (\ref{gapeqdelall}) 
in Section IV.  It will turn out to be close to, but
not identical to, that derived in Ref. \cite{BowersLOFF}.
The difference is that here we have kept the contributions of
particles, holes, and antiparticles in the gap equation,
whereas in Ref. \cite{BowersLOFF} only particle-particle
and hole-hole pairing was considered. 
Pairing in the crystalline
color superconducting phase is dominated by those
pairs in which both particles or both holes in a pair are near their
respective Fermi surfaces. Indeed, it is the fact that such
pairs exist even at nonzero $\delta\mu$ as long as 
$|{\bf q}|\geq\delta\mu$
which explains why the crystalline color superconducting
phase may be favored in the first place.  
We therefore expect that neglecting
the contributions of the antiparticles in the gap
equation, as was done in Ref. \cite{BowersLOFF}, 
should be a good approximation.  Demonstrating this
requires complicating the gap equation considerably at first,
although it does eventually simplify as will be shown by the end of
Section III. We shall see in Section IV that once the
contributions of antiparticles have been
eliminated, the gap equation we derive here
agrees with that of Ref. \cite{BowersLOFF}.

\section{Eliminating Antiparticles}

In order to eliminate the (small) contribution of
the antiparticles on the right-hand side of the gap
equation, we shall need the projectors \cite{PisarskiRischke,SW3}
\begin{eqnarray} \label{projectors}
P_+(p) = {{1+\vec\alpha\cdot\hat p}\over 2} \nonumber \\
P_-(p) = {{1-\vec\alpha\cdot\hat p}\over 2}
\end{eqnarray}
with $\vec\alpha=\gamma_0\vec\gamma$,
where $P_+$ projects onto particle states and $P_-$ projects onto
antiparticle states.  This allows us to separate the ansatz for 
${\bf \Delta}$ into those parts which include antiparticle
pairing and those which do not. 
We replace the ansatz (\ref{simpleansatz}) by 
\begin{eqnarray} \label{gapmat}
{\bf \Delta}^{\alpha\beta}(p,-q) = \epsilon^{\alpha\beta 3} C\gamma^5 &
[ {\Delta_1}P_+(p-q)P_+(p+q) + {\Delta_2}P_-(p-q)P_-(p+q) \nonumber \\
& + {\Delta_3}P_+(p-q)P_-(p+q) + {\Delta_4}P_-(p-q)P_+(p+q) ]\ .
\end{eqnarray}
Here, $\Delta_{1,2,3,4}$ are four (potentially different) gap parameters
whose meaning we now explain. To understand each of the terms in
(\ref{gapmat}), note that, for example,
$$
C \gamma^5 P_+(p-q) P_+(p+q) = P_+^T(-p+q) C \gamma^5 P_+(p+q)\ .
$$
Thus, $\Delta_1$ describes pairing between particles (and not
antiparticles) with momenta ${\bf p}+{\bf q}$ and $-{\bf p}+{\bf q}$.
Similarly, $\Delta_2$ describes antiparticle--antiparticle pairing, and
$\Delta_3$ and $\Delta_4$ describe particle--antiparticle pairing,
which is only possible for ${\bf q}\neq 0$.  (For ${\bf q}=0$, 
the only projectors which occur are $P_+(p)$ and $P_-(p)$, and
$P_+(p)P_-(p)=0$.)

With our point-like interaction, ${\bf \Delta}$ must be independent
of $p$. This requires
\begin{equation}\label{allDeltasequal}
\Delta_1=\Delta_2=\Delta_3=\Delta_4\equiv \Delta\ ,
\end{equation}
which restores the simple ansatz (\ref{simpleansatz}).
It may seem perverse, but we now derive
coupled gap equations for $\Delta_{1,2,3,4}$, without assuming
that they are equal.  We do so for two reasons. First, in a future 
publication, we plan to
restore the gluon propagator. In this context, ${\bf \Delta}$ is
not independent of $p$ and $\Delta_{1,2,3,4}$ therefore
need not all be the same.  The exercise below therefore lays the groundwork
for this future calculation. Second, and in the present context,
we wish to eliminate all terms on the right-hand side of
the gap equation which depend on $\Delta_{2,3,4}$, as they
make only a small contribution.
The reader not interested
in details of this derivation can safely skip to 
Eqs. (\ref{GapEq}) and (\ref{GapDelta}) and the discussion
that follows them.

The Schwinger-Dyson equation (\ref{SDeq}) (using \eqref{pointlike})
is an equation for the matrix ${\bf \Delta}$:
\begin{equation}\label{MMprime}
{\bf \Delta}(k,q) = 3 i G \int \frac{d^4p}{(2\pi)^4} \left(\gamma_\mu
\frac{\lambda^A}{2}\right)^T S_{42}(p,q) \left(\gamma^\mu
\frac{\lambda^A}{2}\right) 
\end{equation}
where $S_{42}(k,q)$ is the (4,2)-component of the fermion propagator
found by inverting \eqref{Sinv}. 
After
some algebra, we find 
\begin{equation} \label{S42def}
S_{42}(p,q) = -{\left(\slash{p}+\slash{q}-\mu_u\gamma^0\right)^{-1}}^T
{\bf \Delta}(p,q) \left[\slash{p}-\slash{q}+\mu_d\gamma^0 - \bar{{\bf
\Delta}}(p,q) {\left(\slash{p}+\slash{q}-\mu_u\gamma^0\right)^{-1}}^T
{\bf \Delta}(p,q)\right]^{-1} \ .
\end{equation}
Upon inserting the ansatz (\ref{gapmat}) for  ${\bf \Delta}(p,q)$,
we can rewrite (\ref{S42def}) in terms of the gap  
parameters $\Delta_{1,2,3,4}$. In order to display the 
resulting expressions, we must first define:
\begin{eqnarray} \label{ABC}
A &=& p_0 + \mu_d - |{\bf p}-{\bf q}| - \Delta_1\frac{\Delta_1
\sin^2(\beta/2) + 1/2 \Delta_3 \cos\beta}{p_0 - \mu_u +
|{\bf p}+{\bf q}|} - \Delta_4\frac{\Delta_4 \cos^2(\beta/2) - 1/2
 \Delta_2 \cos\beta}{p_0 - \mu_u - |{\bf p}+{\bf q}|} \nonumber \\
B &=& p_0 + \mu_d + |{\bf p}-{\bf q}| - \Delta_2\frac{\Delta_2
\sin^2(\beta/2) + 1/2 \Delta_4 \cos\beta}{p_0 - \mu_u -
|{\bf p}+{\bf q}|} - \Delta_3\frac{\Delta_3 \cos^2(\beta/2) - 1/2
 \Delta_1 \cos\beta}{p_0 - \mu_u + |{\bf p}+{\bf q}|} \\
C &=& -\frac{1}{2} \left( \frac{\Delta_1\Delta_3}{p_0 - \mu_u +
 |{\bf p}+{\bf q}|} - \frac{\Delta_2\Delta_4}{p_0 - \mu_u -
 |{\bf p}+{\bf q}|}\right)\ . \nonumber
\end{eqnarray}
Here, 
the angle $\beta$ is defined as the angle between the up quark momentum
${\bf q}+{\bf p}$ and the down quark momentum ${\bf q}-{\bf p}$
and is therefore given by
\begin{equation} \label{beta}
\cos\beta = \widehat{({\bf q}+{\bf p})}\cdot\widehat{({\bf q}-{\bf p})} \ .
\end{equation}
With these definitions, $S_{42}(p,q)$ becomes
\begin{equation}
S_{42}(p,q)=\epsilon^{\alpha\beta 3}C\gamma_5 T(p,q)
\end{equation}
with
\begin{eqnarray} \label{S42}
T(p,q)\!=\!-\!\!&\left[\!\frac{B}{AB-C^2+(A-B)C\cos\beta}\!\!\right. 
\!& \left( \frac{\Delta_1}{p_0-\mu_u+|{\bf p}+{\bf q}|} P_-(p+q)P_-(p-q)
\right. \nonumber \\
&& + \left. \frac{\Delta_4}{p_0-\mu_u-|{\bf p}+{\bf q}|}
P_+(p+q)P_-(p-q) \right) \nonumber \\
&+\!\frac{A}{AB-C^2+(A-B)C\cos\beta}\!&\left( 
\frac{\Delta_3}{p_0-\mu_u+|{\bf p}+{\bf q}|} P_-(p+q)P_+(p-q)
\right. \nonumber \\
&&+ \left. \frac{\Delta_2}{p_0-\mu_u-|{\bf p}+{\bf q}|}
P_+(p+q)P_+(p-q) \right) \\ 
&+\!\frac{C}{AB-C^2+(A-B)C\cos\beta}\!& \left( \frac{\Delta_3 -
\Delta_1}{p_0-\mu_u+|{\bf p}+{\bf q}|} P_-(p+q)P_+(p-q)P_+(p+q) 
\right. \nonumber \\
&&+ \frac{\Delta_4 - \Delta_2}{p_0-\mu_u-|{\bf p}+{\bf q}|}
P_+(p+q)P_+(p-q)P_-(p+q) \nonumber \\
&& \!\left.\!\left.\!-\frac{\Delta_1\!\sin^2\!\frac{\beta}{2}\!+\!\Delta_3\!
\cos^2\!\frac{\beta}{2}}{p_0-\mu_u+|{\bf p}+{\bf q}|}\!P_-(p+q)\!+\!
\frac{\Delta_2\!\sin^2\!\frac{\beta}{2}\!+\!\Delta_4\!
\cos^2\!\frac{\beta}{2}}{p_0-\mu_u-|{\bf p}+{\bf q}|}\!P_+(p+q)
\right)\right] \ .\nonumber 
\end{eqnarray}
Noting that 
$$ 
\left(\frac{\lambda^A}{2}\right)^T \lambda^2
\left(\frac{\lambda^A}{2}\right) = - \frac{2}{3} \lambda^2 \ ,
$$
we 
obtain the gap equation
\begin{eqnarray} \label{GapEq}
&{\Delta_1}P_+(k+q)P_+(k-q) + {\Delta_2}P_-(k+q)P_-(k-q) &\nonumber \\
+&{\Delta_3}P_+(k+q)P_-(k-q) + {\Delta_4}P_-(k+q)P_+(k-q) &= -2 i G\!\int\!
\frac{d^4p}{(2\pi)^4} \gamma_\mu T(p,q) \gamma^\mu\ .
\end{eqnarray}
Upon setting all the $\Delta$'s equal as in (\ref{allDeltasequal}),
the gap equation (\ref{GapEq}) yields
\begin{equation} \label{GapDelta}
\Delta = - 2 i G \int \frac{d^4p}{(2\pi)^4} {\rm Tr }\, T(p,q)\ .
\end{equation}
When written explicitly, this is 
Eq. (\ref{gapeqdelall}).

Using the definition of $T(p,q)$, we can easily identify the
contributions of the four $\Delta$'s to the right-hand side of the gap
equation. Since we expect that the $\Delta_1$ terms, which describe
particle-particle and hole-hole pairing, will give the dominant
contributions to the gap integral, we now eliminate all terms in
$T(p,q)$  which depend on $\Delta_{2,3,4}$.  This means that the
contribution of the antiparticles to the right-hand side of the gap
equation has been eliminated. Note that we are {\it not} setting
$\Delta_{2,3,4}=0$. With a point-like interaction, all the 
$\Delta$'s are in fact equal as in (\ref{allDeltasequal}),
and the left-hand side of the gap equation (\ref{GapEq}) is
$k$-independent and equal to $\Delta$.
The point is that the contributions of those terms 
on the right-hand side of (\ref{GapEq}) in which
$\Delta_{2,3,4}$ appear must be small, and we can
therefore neglect them.  In other words, once all integrations
have been completed on the right-hand side, we would find
that $\Delta_{2,3,4}$ only occur multiplied by quantities which
are small if $\Delta/\mu$ is small.
Dropping $\Delta_{2,3,4}$ on the right-hand side before
integration (but keeping them on the left-hand side) 
should therefore be a good approximation.
The resulting gap equation can be written (in Euclidean space) as
\begin{equation} \label{gapeqdel1}
\Delta = 2 G \int \frac{d^4p}{(2\pi)^4} \frac{2 \Delta
\sin^2\frac{\beta}{2}}{\left(p_0-i E_1({\bf p})\right)
\left(p_0+i E_2({\bf p})\right)} 
\end{equation}
where $E_{1,2}({\bf p})$ are defined as in Ref. \cite{BowersLOFF}:
\begin{eqnarray} \label{E12}
E_1({\bf p}) = & + \delta\mu + \frac{1}{2}
\left(|{\bf p}+{\bf q}|-|{\bf p}-{\bf q}|\right) + \frac{1}{2}
\sqrt{\left(|{\bf p}+{\bf q}|+|{\bf p}-{\bf q}|-2\bar{\mu}\right)^2 +
4 \Delta^2 \sin^2\frac{\beta}{2}} \nonumber \\
E_2({\bf p}) = & -\delta\mu - \frac{1}{2}
\left(|{\bf p}+{\bf q}|-|{\bf p}-{\bf q}|\right) + \frac{1}{2}
\sqrt{\left(|{\bf p}+{\bf q}|+|{\bf p}-{\bf q}|-2\bar{\mu}\right)^2 +
4 \Delta^2 \sin^2\frac{\beta}{2}}
\end{eqnarray}
and $\beta$ is defined in \eqref{beta}. As we describe below,
we have confirmed explicitly that the gap equation (\ref{gapeqdel1})
is a  good approximation to Eq. (\ref{gapeqdelall}).

\section{Analyzing the Gap Equation}

The energies $E_1$ and $E_2$ given in (\ref{E12})
arise in Ref. \cite{BowersLOFF}.  There, we deduced that
the right-hand side of the gap equation
must be taken to vanish in those regions of ${\bf p}$-space where
either $E_1({\bf p})$ or $E_2({\bf p})$ is negative via the
following argument.
In the region where $E_1({\bf p})<0$, it is free-energetically
favorable to have unpaired $u$-quarks rather than pairs. 
Similarly, in the region where $E_2({\bf p})<0$, it is free-energetically
favorable to have unpaired $d$-quarks rather than pairs. 
Because quarks do not pair in these
 ``blocking regions'' of momentum space, these regions
do not contribute to the gap equation, which becomes an integral
over those regions of momentum space wherein pairing occurs.
If $\Delta$ is set to zero, the blocking regions 
are simply described. They are
the regions in ${\bf p}$-space where the 
$u$-quark state with momentum ${\bf p}+{\bf q}$ is within
the $u$ Fermi sea while the
$d$-quark state with momentum ${\bf -p}+{\bf q}$ is
outside the $d$ Fermi sea, or vice versa.  
In the presence of 
a nonzero $\Delta$, the boundaries of the blocking regions
are given by $E_1({\bf p})=0$ and $E_2({\bf p})=0$ and therefore
depend on $\Delta$ and are not simply determined by the 
locations of the noninteracting
Fermi surfaces.
The result of this analysis, presented in Ref. \cite{BowersLOFF},
is a variational procedure in which
the boundaries of the blocking regions, and thus
the specification of the variational ansatz itself, 
depend on the gap $\Delta$,
which is in turn obtained by solving a gap equation whose
integrand is restricted by hand to vanish within said
blocking regions.

In contrast to the intricacy of the variational approach,
the physics of the blocking regions
emerges from a completely straightforward analysis of
the gap equation 
in the form we have derived above, namely Eq. (\ref{gapeqdel1}).
We simply do the $p_0$ integral by contour integration. 
There are two poles, both of which lie on the imaginary axis.
Let us close the contour in the upper half plane.
If $E_1>0$ and $E_2>0$, we pick up the pole at $p_0=iE_1$ which
has a residue proportional to $1/(E_1+E_2)$.  If $E_1<0$ and 
$E_2>0$, both poles are in the lower half plane, and the 
right-hand side of the gap equation vanishes.  If 
$E_1>0$ and  $E_2<0$, both poles are in the upper half plane,
the residues from the two poles cancel, and 
the right-hand side of the gap equation again vanishes. 
(If we close the contour in the lower half plane,
we obtain the same result upon noticing that
we encircle no poles if $E_1<0$ and 
$E_2>0$ and two poles with cancelling residues for
$E_1>0$ and  $E_2<0$.)
Thus, upon
doing the $p_0$ integration we obtain the gap equation of
Ref. \cite{BowersLOFF}: 
\begin{eqnarray} \label{ABRgapeq}
1 &=& 2 G \int_{{\bf p} \in {\cal P}}  \frac{d^3p}{(2\pi)^3} 
\frac{2\sin^2\frac{\beta}{2}}{E_1({\bf p})
+E_2({\bf p})}\nonumber\\
&=& 2 G \int_{{\bf p} \in {\cal P}}  \frac{d^3p}{(2\pi)^3} \frac{2
\sin^2\frac{\beta}{2}}{\sqrt{\left(|{\bf p}+{\bf q}| +
|{\bf p}-{\bf q}|-2\bar{\mu}\right)^2 + 4 \Delta^2
\sin^2\frac{\beta}{2}}}
\end{eqnarray}
where the ``pairing region'' ${\cal P}$ in ${\bf p}$-space is
given by
\begin{equation} \label{Preg}
{\cal P} = \{ {\bf p} \ | \ E_1({\bf p}) > 0 \ {\rm and} \ E_2({\bf p}) >
0 \} \ .
\end{equation}
Thus, a trivial exercise in residue calculus has reproduced
the blocking regions, excluding from the gap equation
those 
regions in momentum
space where $E_1({\bf p})$ or $E_2({\bf p})$ is negative.
Note 
that because $E_1({\bf p}) + E_2({\bf p})\geq 0$,
as can be seen from the definitions (\ref{E12}), there is no value of
${\bf p}$ for which both $E_1$ and $E_2$ are negative.
Note also that the 
gap equation is dominated by those regions in momentum
space where $E_1({\bf p})+E_2({\bf p})$ is as small
as possible, where the integrand in (\ref{ABRgapeq}) 
is of order $1/\Delta$.
These values of ${\bf p}$ are such that both members
of a LOFF pair have momenta close to (within $\sim \Delta$ of) 
their respective Fermi surfaces. That is, $|{\bf p+q}|$ is within
$\Delta$ of $\mu_u$ and $|{\bf -p+q}|$ is within $\Delta$ of $\mu_d$.

For completeness, we sketch the analysis of 
(\ref{gapeqdelall}), in which the contributions of antiparticle 
pairing to the gap equation
have not been eliminated. The denominator of the integrand 
in (\ref{gapeqdelall}) is a fourth order polynomial in $p_0$, so the
gap equation can be rewritten as 
\begin{equation} \label{gapeqallP}
\Delta = 2 G \int \frac{d^4p}{(2\pi)^4} \frac{4 \Delta w}{\left(
p_0 -i P_1({\bf p}) \right) \left( p_0 +i P_2({\bf p}) \right) \left(
p_0 +i \bar{P_1}({\bf p}) \right) \left( p_0 -i \bar{P_2}({\bf p}) 
\right)}\ .
\end{equation}
The analytical expressions for the poles 
$P_1({\bf p})$, $P_2({\bf p})$,
$\bar{P_1}({\bf p})$, and $\bar{P_2}({\bf p})$ are complicated
and uninformative.  However, we have checked 
that for reasonable choices of parameters, 
the numerical values of $P_1({\bf p})$ and $P_2({\bf p})$
are very close to $E_1({\bf p})$ and $E_2({\bf p})$
and those of $\bar{P_1}({\bf p})$ and $\bar{P_2}({\bf p})$
are very close to the antiparticle energies
\begin{eqnarray} \label{Ebar12}
\bar{E_1}({\bf p}) = & - \delta\mu + \frac{1}{2}
\left(|{\bf p}+{\bf q}|-|{\bf p}-{\bf q}|\right) + \frac{1}{2}
\sqrt{\left(|{\bf p}+{\bf q}|+|{\bf p}-{\bf q}|+2\bar{\mu}\right)^2 +
4 \Delta^2 \sin^2\frac{\beta}{2}} \nonumber \\
\bar{E_2}({\bf p}) = & + \delta\mu - \frac{1}{2}
\left(|{\bf p}+{\bf q}|-|{\bf p}-{\bf q}|\right) + \frac{1}{2}
\sqrt{\left(|{\bf p}+{\bf q}|+|{\bf p}-{\bf q}|+2\bar{\mu}\right)^2 +
4 \Delta^2 \sin^2\frac{\beta}{2}} \ .
\end{eqnarray}
The analysis of (\ref{gapeqallP}) is analogous to that of 
(\ref{gapeqdel1}).  Wherever both $P_1$ and $P_2$ are positive,
the pole at $p_0=i P_1$ contributes, 
with residue proportional to $1/(P_1+P_2)$.
The integral is dominated by the region where $P_1+P_2$ is 
close to zero. There is also a contribution from the pole at 
$p_0=i \bar P_2$, but the residue of this pole is 
nowhere large.
For this reason, and because $P_1$ and $P_2$ are 
numerically very close to $E_1$ and $E_2$, we find
that (\ref{gapeqdel1}), and thus
(\ref{ABRgapeq}) which was derived 
variationally in Ref. \cite{BowersLOFF}, 
is a good approximation to (\ref{gapeqallP}). 
For
$\bar\mu=400$~MeV and $G$ chosen such that the BCS gap at $\delta\mu=0$
is $100$~MeV, we take $\delta\mu$ within the range where
the crystalline color superconducting phase is favored \cite{BowersLOFF}
and choose a value of $\Delta$ which solves (\ref{gapeqdel1}). We
then find that the right-hand sides of (\ref{gapeqdel1}) and
(\ref{gapeqallP}) differ by about 20\%.  The discrepancy
vanishes in the weak-coupling limit.
(Note
that eliminating the contribution of the pole at 
$p_0=i \bar P_2$ does not by itself reduce (\ref{gapeqallP}) to 
(\ref{gapeqdel1}), although it does change the 20\%
discrepancy to a 1\% discrepancy.
Eliminating all the contributions of 
the antiparticles is more subtle, as we have seen.)

Although the expressions for the $P({\bf p})$'s are much more
complicated than those for the $E({\bf p})$'s, the gap equation
(\ref{gapeqallP}) actually turns out to be more easily solvable (by
Mathematica) than (\ref{gapeqdel1}), as we now explain.  
The blocking regions in (\ref{gapeqallP})  are
regions wherein either $P_1$ or $P_2$ is negative, and are therefore
bounded by surfaces on which $P_1$ or $P_2$ vanishes. Within these
regions, there is no contribution from the $P_1$ and $P_2$
poles.
The simplification which occurs in
(\ref{gapeqallP}) is in the explicit expressions for 
the boundaries of the blocking regions. 
We simply set $p_0=0$ in the denominator of 
Eq. (\ref{gapeqdelall}) and for each value of 
$|{\bf p}|$ we solve for $\cos\theta$, the angle between
${\bf p}$ and ${\bf q}$:
\begin{equation} \label{xbound}
\cos\theta = - \frac{2 \bar\mu \delta\mu \pm \sqrt{
\left( |{\bf p}|^2 - |{\bf q}|^2 -
\delta\mu^2 + \bar{\mu}^2 + \Delta^2 \right)^2 - 4 (|{\bf p}|^2 - \delta\mu^2)
(\bar{\mu}^2 - |{\bf q}|^2)}}{2 |{\bf p}||{\bf  q}|} \ .
\end{equation}
This allows one to implement the fact that the 
${\bf p}$-integral is to be taken over all of ${\bf p}$-space except for
the blocking regions via explicitly specified
limits on the $\cos\theta$ integral.
In contrast, even though $E_1$ and $E_2$ are simpler than
$P_1$ and $P_2$,  the dependence of the denominator
in (\ref{gapeqdel1}) on $\cos\theta$ is 
more complicated than that in (\ref{gapeqdelall}),
and the blocking regions can only be specified explicitly
as roots of a quartic polynomial.

The calculation of $\delta\mu_2$  
of Ref. \cite{BowersLOFF} follows directly from 
Eq. (\ref{gapeqdel1}), as does the 
value of $|{\bf q}|$, which is given to a very good
approximation by that for
$\delta\mu\rightarrow\delta\mu_2$.  
(One finds $\delta\mu_2$ by seeking the largest value of $\delta\mu$ for
which there is a choice of $|\bf q|$ which yields a nonzero 
solution $\Delta$ to the gap equation (\ref{gapeqdel1}) \cite{BowersLOFF}.)
Our rederivation of (\ref{gapeqdel1})
has several merits.
First, as it begins with a Schwinger-Dyson equation rather
than a variational wave function, it may appear more familiar. 
Second, the emergence of blocking
regions is straightforward.
Third, it 
is the basis for many generalizations:  Quark masses
can easily be introduced in (\ref{Sinv}), and the analogue
of (\ref{gapeqdelall}) can then be derived.  
The gluon propagator need not be replaced by a point-like
interaction. This opens the way to a treatment of color superconductivity
at asymptotically high density.  Also, nontrivial crystal
structures can be analyzed beginning with a Nambu-Gorkov propagator
which admits ``anomalous propagation'' in which $2{\bf q}$ 
of momentum is gained or lost, for several different values of the 
vector $2{\bf q}$.  Finally, the generalization to nonzero
temperature is straightforward. To this we now turn.

\section{Crystalline Color Superconductivity at Nonzero Temperature}

We can now derive the gap equation for the LOFF state at nonzero
temperature. We begin with \eqref{gapeqdel1}.
Using the standard formalism, we
obtain the nonzero temperature gap equation by
converting the $p_0$ integral into a sum over
Matsubara frequencies.  That is, with $\omega_n = (2n+1)\pi T$, using the
prescription 
$$p_0 \rightarrow \omega_n \ \ {\rm and} \ \ \int\frac{d p_0}{2\pi}
\rightarrow T \sum_n$$
we obtain the following equation:
\begin{equation} \label{GapEqMatsubara}
\Delta = 2 G \int\frac{d^3p}{(2\pi)^3} 2 \Delta
\sin^2\left(\frac{\beta}{2}\right) \; T \sum_{n=-\infty}^{\infty}
\frac{1}{\left( \omega_n-iE_1({\bf p})\right) \left( \omega_n +
iE_2({\bf p}) \right)} \ .
\end{equation}
The sum may be evaluated by converting it into a contour integral:
$$\sum_{n=-\infty}^{\infty} \frac{1}{(\omega_n-iE_1)(\omega_n
+iE_2)} = \frac{1}{2 \pi i} \int_C dz \frac{1}{T(e^{z/T}+1)} \;
\frac{1}{(z + E_1)(z - E_2)}$$ 
where the contour $C$ encircles the imaginary axis.\footnote{Note that
$\frac{-1}{T(e^{z/T}+1)}$ has simple poles with unit residue  at $z = i
\omega_n$.}  We may deform the contour so that it encircles, with
negative orientation, the poles off the imaginary axis.  This gives 
\begin{eqnarray*}
 \frac{1}{2 \pi i} \int_C dz \frac{1}{T(e^{z/T}+1)} \frac{1}{(z +
E_1)(z - E_2)} &=& -\frac{1}{T(e^{E_2 /T}+1)} \frac{1}{E_2 + E_1} +
\frac{1}{T(e^{-E_1 /T}+1)} \frac{1}{E_2 + E_1} \\ 
&=& \frac{1}{2T(E_1+E_2)} \left[\tanh\left(\frac{E_1}{2T}\right) +
\tanh\left(\frac{E_2}{2T}\right)\right] \ .
\end{eqnarray*}
Upon using these identities in the finite-temperature gap
equation \eqref{GapEqMatsubara}, we obtain:
\begin{equation} \label{GapEqT}
1 = 2 G \int\frac{d^3p}{(2\pi)^3} \frac{2
\sin^2\left(\frac{\beta}{2}\right)}{E_1({\bf p}) + E_2({\bf p})} \ \frac{1}{2}
\left[\tanh\left(\frac{E_1({\bf p})}{2T}\right) +
\tanh\left(\frac{E_2({\bf p})}{2T}\right)\right] \ .
\end{equation}
Note that here the integration is performed over all of
${\bf p}$-space: there are no blocking regions at nonzero temperature.  
The blocking regions
emerge in the limit $T \rightarrow 0$ as follows: if 
$E_1({\bf p}) > 0$ and $E_2({\bf p}) < 0$, then $\tanh(\frac{E_1({\bf p})}{2T})
\rightarrow 1$ while $\tanh(\frac{E_2({\bf p})}{2T}) \rightarrow -1$,
and the integrand vanishes. The same result holds if $E_1({\bf p}) <
0$ and $E_2({\bf p}) > 0$.  If, however, $E_1({\bf p}) > 0$ and
$E_2({\bf p}) > 0$,
$$\frac{1}{2} \left[\tanh\left(\frac{E_1({\bf p})}{2T}\right) +
\tanh\left(\frac{E_2({\bf p})}{2T}\right)\right] \rightarrow 1$$
reproducing the zero temperature gap equation \eqref{ABRgapeq}. 
Note that even if our goal were
just to understand physics at $T=0$ it may be of practical
value to do calculations at several nonzero values of the
temperature and then extrapolate to $T=0$.  The reason is
that at $T=0$, specifying the boundaries of ${\cal P}$ of
(\ref{Preg}), and thus the limits of integration, can 
be a numerical challenge. At any nonzero temperature, instead,
no limits of integration need be specified. The $\tanh$ factors
impose the required limits as $T$ gets small.

We shall present $T\neq0$ results for parameters chosen
as in Fig. 4 of Ref. \cite{BowersLOFF}, which we
first recapitulate. We
specify the four-fermion
interaction by choosing the cutoff parameter,
defined in Ref. \cite{BowersLOFF}, to be
$\Lambda=1$~GeV and 
requiring $G$ to be such that the BCS gap is $\Delta_0=40$~MeV
at $\delta\mu=0$. We choose $\bar\mu=400$~MeV, and explore
different values of $\delta\mu$.   
At $T=0$\cite{BowersLOFF}, we find nonzero solutions to the 
gap equation 
(\ref{ABRgapeq}) for $\delta\mu<\delta\mu_2=0.744\Delta_0$.
Above $\delta\mu_2$,
no pairing between $u$ and $d$ quarks is possible. 
The crystalline color superconductor phase  has
lower free energy than the ordinary BCS phase as long
as $\delta\mu>\delta\mu_1=0.710\Delta_0$, where a first order phase
transition occurs.  
A precise determination of
$\delta\mu_1$ requires expressions for the
free energy of both phases. 
The free energy of the crystalline color
superconductor phase could be obtained from the gap equation 
along the lines described in Section 4.3 of Ref. \cite{ioffe}.
However, 
$\delta\mu_1$ is well approximated by the
$\delta\mu$ at which the BCS and unpaired states have
equal free energy, which turns out to be $0.711\Delta_0$ \cite{BowersLOFF}.
At $\delta\mu=\delta\mu_1$, $\Delta=7.8$~MeV \cite{BowersLOFF}. 
For $\delta\mu\rightarrow\delta\mu_2$ from below, 
$\Delta$ vanishes like $(\delta\mu_2-\delta\mu)^{1/2}$ \cite{BowersLOFF}.  
The window
$\delta\mu_1<\delta\mu<\delta\mu_2$ widens if the
interaction includes attraction in the spin-one channel.
We expect this window to widen at asymptotic density,
where quarks interact by exchanging a propagating gluon.

\begin{figure}[t]
\centering
\epsfig{file=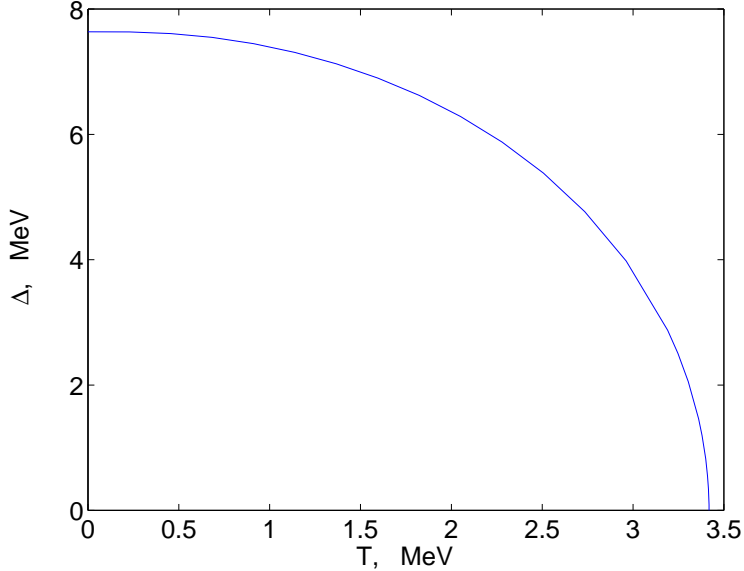,width=3.9in}
\vspace{0.2in}
\caption{The gap $\Delta$ as a function of temperature $T$,
at $\delta\mu=\delta\mu_1$.  
At zero temperature, $\Delta=7.8$~MeV$=0.195\Delta_0$. 
The gap vanishes above $T_c=3.42$~MeV.}
\vspace{0.2in}
\end{figure}
\begin{figure}[t]
\centering
\qquad
\epsfig{file=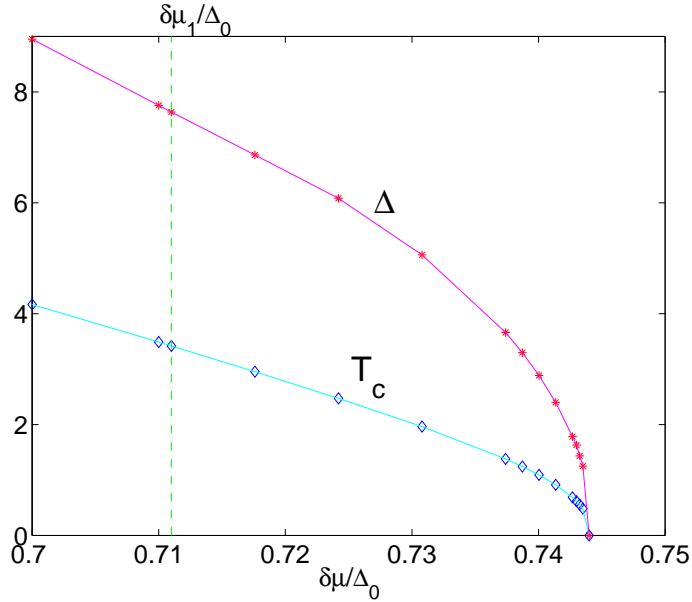,width=3.6in}
\vspace{0.2in}
\caption{The zero temperature gap $\Delta$ and 
the critical temperature $T_c$ (both in MeV) 
as functions of $\delta\mu$.  $\Delta$ vanishes
like $(\delta\mu_2-\delta\mu)^{1/2}$ for $\delta\mu\rightarrow\delta\mu_2$.
In this limit, $T_c/\Delta\rightarrow 0.39$.  To the left of
$\delta\mu_1$, the ordinary BCS phase is favored.}
\end{figure}
In Fig. 1, we show the dependence of $\Delta$ on the temperature $T$
at $\delta\mu=\delta\mu_1$.  We find that the critical temperature
above which the crystalline color superconductivity
is lost is $T_c=3.42$~MeV, corresponding to $T_c=0.44 \Delta(T=0)$.
In Fig. 2, we plot both $T_c$ and $\Delta(T=0)$ as functions
of $\delta\mu$, for $\delta\mu_1<\delta\mu<\delta\mu_2$.
We find that the ratio $T_c/\Delta(T=0)$ changes little, decreasing from
0.44 at $\delta\mu_1$ to $0.39$ for $\delta\mu\rightarrow\delta\mu_2$.
This agrees with the previously known result that 
$T_c/\Delta(T=0)\rightarrow\sqrt{3/2\pi^2}$
for $\delta\mu\rightarrow\delta\mu_2$ \cite{Takada2}.
There are two {\it a priori}
reasons why one may have questioned whether $T_c$ 
in the crystalline color superconducting phase would
turn out to be proportional
to $\Delta(T=0)$. 
First, this phase is in fact gapless. There are directions in momentum
space (which intersect the boundaries of the
blocking regions) for which gapless excitations exist at zero temperature.
One may wonder whether the presence
of these gapless modes, which can be excited at arbitrarily
low temperature, might lower $T_c$.  Second, the condensation
energy in the crystalline color superconductor phase is 
of order $\bar\mu^2\Delta^4/\Delta_0^2$ \cite{BowersLOFF},
whereas that in the ordinary BCS phase is of order $\bar\mu^2\Delta_0^2$.
One may therefore wonder whether the $T_c$ 
for crystalline color superconductivity scales differently
with $\Delta$.  It turns out, however, that the
simplest expectation holds true:  $\Delta$ is the gap
in the fermion spectrum in directions in momentum space along
which pairing is maximized and destroying the condensate therefore requires
a temperature $T_c$ which is of order $\Delta$.

\acknowledgments

We acknowledge helpful discussions with M. Alford.

Research supported in part  by the U.S. Department
of Energy (D.O.E.) under cooperative research agreement
DE-FC02-94ER40818. The work of KR is supported in part by a DOE OJI
grant and by the Alfred P. Sloan Foundation. The work of JAB 
and JK is
supported in part by DOD National Defense Science and Engineering
Graduate Fellowships.

\end{document}